\documentclass[pre,preprint,showpacs,showkeys]{revtex4}
\usepackage{bm}
\usepackage{times}
\usepackage{amsmath}
\usepackage{amsfonts,amssymb}
\usepackage{natbib}

\newcommand{\pd}[2]{\frac{\partial #1}{\partial #2}}

\usepackage{verbatim}
\usepackage{graphicx}
\usepackage{color}
\newcommand{\hb}[1]{\hat{\bm{#1}}}
\newcommand{\jgr}{J. Geophys. Res.}

\newcommand{\apjl}{Astrophys. J. Lett.}
\newcommand{\grl}{Geophys. Res. Lett.}
\allowdisplaybreaks
\begin{document}
\title{Efficient electron heating in relativistic shocks and gamma ray burst afterglow}
\author{M. Gedalin$^1$, M. A. Balikhin$^2$,  and D. Eichler$^1$}
\affiliation{$^1$Department of Physics, Ben-Gurion University, Beer-Sheva, Israel\\ $^2$ACSE, University of Sheffield, Sheffield, UK}

\begin{abstract}
Electrons in shocks are efficiently energized due to the cross-shock
potential, which develops because of differential deflection of
electrons and ions by the magnetic field in the shock  front. The
electron  energization is necessarily accompanied by scattering and
thermalization. The mechanism is efficient in both magnetized and
non-magnetized relativistic electron-ion shocks. It is proposed that
the synchrotron emission from the heated electrons in a layer of
strongly enhanced magnetic field is responsible for gamma ray burst
afterglows.
\end{abstract}

\maketitle
\section{Introduction}
Electron energization is usually considered as a secondary problem
at heliospheric shocks, where most attention is paid to ion heating
and reflection. In astrophysical shocks, however, these energized
electrons emit the observed radiation, and are frequently  the only
source of information about the remote astrophysical process. Gamma
ray burst (GRB) afterglow is believed to be  synchrotron emission
from electrons accelerated in the shock that develop during the
interaction of the expanding ultra-relativistic plasma into the
interstellar medium (ISM) \citep{reviews}. Estimates (e.g.,
Ref.~\onlinecite{estimates} and references therein) suggest  that
the  required average energies of electrons reach a sizable part of
the relativistic ion energy, and that the magnetic field in the
emission region should be highly amplified, however, the origins of
the electron heating and the magnetic field amplification remain
poorly understood. In this paper we propose a single mechanism that
accomplishes both, and is driven by the preferential deflection of
electrons versus ions, when the former are lighter than the latter,
by a local increase in the magnetic field.

The mechanism of electron heating in heliospheric shocks is widely
understood as follows \citep{feldman}: Electrons are decelerated more
easily than ions, either by growing coherent magnetic fields in
quasi-perpendicular shocks or by small scale magnetic structures in
quasi-parallel shocks. The developing charge separation, however
small it is, results in the build up of a cross-shock potential
which is a substantial fraction of the incident ion energy. It is
this cross-shock potential which decelerates ions when they become
demagnetized in a thin transition layer of a quasi-perpendicular
shock. In quasi-parallel shocks the parallel component of the
magnetic field does not effect the ion motion along the shock
normal, so that ions effectively become demagnetized just ahead of
the transition. The same cross-shock potential which decelerates
ions should accelerate electrons along the shock normal thus
transferring energy from  ions to electrons. The efficiency of the
process is reduced by the electron drift in the magnetic fields,
 during which they lose energy by drifting down an electric
potential. The final step of the process, electron thermalization,
can be achieved by
turbulent scattering following plasma instabilities.

The mechanism of the prompt electron heating in steady state
magnetized shocks is well-known \citep{mechanism}: electrons become
demagnetized in the shock front if the ramp width is smaller than
their convective gyroradius, or when the cross-shock electrostatic
field becomes sufficiently inhomogeneous to drag them across the
magnetic field. In heliospheric shock these conditions are rarely
satisfied since shocks are rarely this narrow. Moreover, only that
part of the cross-shock potential which cannot be eliminated by
transformation into a de Hoffman-Teller frame  \citep{GS84} can be
effectively used for electron energization.  However, the profiles
become steeper with the increase of the Mach number \citep{ramps} so
that the conditions for demagnetization may be achieved more easily.
The transition layer of quasi-perpendicular non-relativistic shocks
consists of several distinct regions \citep{Sc86}, the steepest
magnetic field increase is a "ramp" (whose width is less than the
ion inertial length $l_i=c/\omega_{pi}$, $\omega_{pi}^2=4\pi
n_ue^2/m_i$) and a large magnetic overshoot (whose width is of the
order of the downstream ion gyroradius). The overshoot height is
found experimentally to increase with the increase of the Mach
number \citep{overshoots}. The ratio of the ramp width to the ion
convective gyroradius $\sim l_i\cos\theta/(V_u/\Omega_u)\sim 1/M$,
where $\theta$ is the angle between the shock normal and the
upstream magnetic field $\Omega_u=eB_u/m_ic$ is the upstream ion
gyrofrequency, and $M=\Omega_u/\omega_{pe}$ is the Alfvenic Mach
number. In perpendicular shocks the ramp width can be as small as
$l_e=c/\omega_{pe}$ \cite{narrow}.

Theory of electron heating in quasi-parallel shocks has been
developed s less elaborately, partly because of the lack of coherent
structure in these shocks. Observations  \citep{1993JGR....98.3875T}
imply that the dominant electron heating process is the same as in
quasi-perpendicular shocks and appear to illustrate the  importance
of the DC effects of the coherent forces for the  physics of
electron heating in shocks.

GRB-generated  forward shocks in the ISM are ultra-relativistic
$\Gamma \ge 20$.  These shocks are parameterized by
$\sigma=B_u^2/4\pi n_um_ic^2\gamma_u\ll1 $ (this is written in the
shock frame but is invariant). They are very high Mach number
shocks, since the corresponding Mach number $M=1/\sigma$. Based on
numerical simulations, it is widely believed that such shocks may
be formed due to the development of Weibel instability
\cite{weibel} into ion current filaments surrounded by regions of
enhanced magnetic field. The filaments are elongated along the
flow direction, with the magnetic field nearly perpendicular to
the shock normal. The  magnetic field around the filaments reaches
nearly equipartition values but the magnetic filling factor is
low. The width of a magnetic region is expected to be of up to
tens of electron inertial length  while the length of the region
over which  the surrounding magnetic field is high is determined
by the ion scale. Although there is no gyration in these
structures  high magnetic fields  at small scales make them play
the role of a perpendicular magnetized shock front in what
concerns electron energization.

 In this paper
we suggest that  differential momentum transfer to ions and
electrons, typical for steady perpendicular shock and filamentary
shock as well, results in the buildup of a strong potential drop,
comparable to the upstream ion energy. The electrons are demagnetized  and receive a
significant fraction of the original ion kinetic energy directly
from the dc electric field. The accelerated electron energy is converted either into gyration energy (by the coherent magnetic field in magnetized shocks) or random motion energy (by small scale magnetic fields in Weibel mediated shocks)   thus resulting in the collisionless heating. In both cases a region of strongly enhanced magnetic field is developed in the shock front, where the heated electrons should efficiently emit synchrotron radiation. We show that although the details of the mechanism differ in magnetized and non-magnetized shocks, the underlying physics is very similar, and the eventual efficiency does not depend on the magnetization.
GRB
afterglows may be explained, at least in part, by radiation from
these heated electrons.

In proposing a mechanism for electron heating based on charge separation we
do not mean to deny the existence of other mechanisms, e.g. decay and merging
of magnetic islands,  which can operate even with equal masses of both species.
However, because the Weibel shock is otherwise required to "wait" for a
bootstrap process in which electrons are heated by magnetic field but magnetic
field growth is limited by electron temperature \citep{LE2006}, we suggest that
in the case of realistic mass ratios even a modest degree of charge separation
can help to jump start the collisionless shock process.

\section{Magnetized shocks.}
As will be seen below  magnetized shocks are more restrictive in producing efficient electron heating, yet the basic features of the mechanism are typical for non-magnetized shocks as well (with suitable modifications). Therefore, we start our analysis with quasi-perpendicular magnetized shocks.

 Relativistic shock propagating obliquely in ISM becomes nearly
perpendicular in the shock frame, because of the Lorentz
transformation, $\theta_{shock}=\theta_{ISM}/\gamma_u\ll 1$ (here
$\gamma_u\gg 1$ is the Lorentz-factor of the shock relative to ISM
or, alternatively, the Lorentz-factor of  the incident plasma flow
in the shock frame). The de Hoffman-Teller frame, which has the
velocity $V_u\tan\theta$ along the shock front, does not exist for
$V_u\approx c$ and $\tan\theta > 1/\gamma_u$. In what follows we
consider first a quasi-stationary perpendicular
magnetized shock front where the fields are given by $B_z=B(x)$,
$E_x(x)$, and $E_y=\text{const}$.

\subsection{Demagnetization conditions.} The condition for
the demagnetization by inhomogeneous $E_x$ is the statement that
the accelerating electric field straightens the trajectory faster
than the magnetic field bends it. The condition can be derived in
the simplest way by approximating the inhomogeneous electric field
with a linear slope while ignoring the magnetic field variations
in the electron equations of motion.
Then the motion is described by $\bm{v}-\bm{v}_0, x-x_0 \propto
\exp(\lambda t)$. Imaginary $\lambda$ ($\lambda^2<0$) corresponds
to the particle gyration in the magnetic field (magnetic bending
prevails) while and $\lambda^2>0$ results in the exponential
acceleration  across the magnetic field, that is, demagnetization
\citep{mechanism}. Relativistic generalization of the calculations
in Ref.~\onlinecite{mechanism} is straightforward (see Appendix~\ref{demagnetization}) and gives
\begin{align}
- \gamma (1+\gamma^2v_y^2/c^2)(e/m_e)\frac{dE_x}{dx}>\Omega_e^2,\label{eq:cond}
\end{align}
where $\Omega_e=eB/m_ec$. If \eqref{eq:cond} is satisfied,
electrons are efficiently accelerated across the magnetic field
and acquire most of the cross-shock potential at the
demagnetization region. The condition is local and cannot be
satisfied in the whole shock transition layer, since $-dE_x/dx>0$
is required. Thus, the electrons can be  demagnetized while
crossing a part of the magnetic inhomogeneity, after which they
may return to be magnetized and the acquired energy is immediately
converted into their gyration energy. Alternatively, electrons become demagnetized if the inhomogeneity scale of the magnetic field $(1/B)(dB/dx)$ is smaller than the convective electron gyroradius $c\gamma_e/\Omega_e$.

The above demagnetization condition is derived in a
simplified assumption that the magnetic field is constant. While
this is not the case inside the shock,
numerical analyses \citep{mechanism} have shown
remarkable agreement with application of the non-relativistic
version of \eqref{eq:cond} at the upstream edge of the ramp,  and
\eqref{eq:cond} should be considered  an estimate.

Demagnetization is required for an electron to utilize the cross-shock potential,
otherwise electrons simply $\bm{E}\times\bm{B}$ drift, and the energy gain due to the potential ($E_x$) is balanced by the energy loss because of the motion along $E_y$. Once the drift is substantially suppressed a net energy gain is achieved \citep{mechanism}. The energy gain is determined by the potential drop across the demagnetization region. When magnetization is restored no further energization occurs. The acquired energy is converted into the electron gyration energy where demagnetization disappears. Further collisionless "randomization" occurs through gyrophase mixing in the nonstationary and inhomogeneous fields of the shock front, thus resulting in the collisionless heating \citep{feldman}. Maxwellization is not required for the existence of the shock.

\subsection{Magnetic structure and cross-shock electric field.}
 For the purpose of description we consider a one-dimensional and steady shock.
 The basic equations of the two-fluid hydrodynamics for this shocks are given in
 Appendix~\ref{basic}. The cross-shock electric field can then be  estimated using
 the momentum conservation:
 \begin{align}
 \sum T_{xx}+\frac{B^2-E^2}{8\pi}=\text{const}, \quad T_{xx}=\langle p_xv_x\rangle. \label{eq:eq}
\end{align}
Here $\langle \ldots \rangle$ means averaging over the distribution function and the summation is over both species.
 The discussion below is based on the basic picture justified by observations \citep{ramps}, simulations \citep{BG89}, and theory \citep{theory},  that the front steepening stops at the width much smaller than the convective ion gyroradius which ensures ion demagnetization inside the shock transition layer, and the assumption that this basic picture applies to relativistic magnetized shocks. As a consequence, ions are only slightly deflected within the transition layer (ramp) while almost all current necessary for the magnetic field increase is produced by electrons, which (partially) experience $\bm{E}\times\bm{B}$ drif. The latter allows one to estimate the electron velocity as
$v_y\sim (c/4\pi n_ue)(dB_z/dx)$.
Before electrons are substantially heated the magnetic force should be balanced by the electric force so that
\begin{align}
E_x\approx -\frac{1}{8\pi n_u}\frac{d}{dx}B_z^2,\Rightarrow
en_u\Delta \phi\approx \Delta B^2/8\pi, \label{eq:phisto}
\end{align}
where we have taken into account approximate quasineutrality and neglected the change of the ion density.
For $\sigma\ll 1$ even slight deceleration of ions causes strong enhancement of the magnetic field, which results in the development of the cross-shock potential which, in turn, further decelerates ions. A spontaneous small enhancement of the upstream magnetic field causes exponential development of the magnetic field increase at the typical electron length scale (see below). The corresponding electric field given by \eqref{eq:phisto}.

Upon crossing this narrow region of the magnetic field increase and potential development ions begin to gyrate.
Assuming the gyrating ions to be a cold beam, it is easy to
see that the momentum flow $T_{xx}$ in the particles is very
small where the ions have gyrated by 90 degrees and are
moving nearly perpendicular to the flow ($x$) direction. If
the shock is to be quasi-stationary, this must be taken up
by some combination of magnetic and electron pressure.
For a weakly magnetized shock, magnetic pressure balance would imply a magnetic field far larger than that dictated by shock jump conditions. Electron pressure would
require significant cross shock potential. The two quantities are connected by equation \eqref{eq:phisto}, so the argument implies
both magnetic overshoot and a large cross-shock potential.
Since the shock may be unsteady, this argument does not
constitute a rigorous proof of either, however, it shows that
ion reflection is likely to cause extremely chaotic conditions in which pressure balance without strong cross shock
potential and magnetic overshoot would seem to require
implausibly
fine
tuning.

In order to know whether electrons are indeed demagnetized one has
to know the spatial profile of the shock. Two fluid hydrodynamics
predicts  \citep{theory} that  a
perpendicular magnetosonic wave steepens down to the slope
determined by the electron inertial length $l_e$. Following the
general principles of
\citep{theory}, we seek nonlinear
wave solutions that are asymptotically homogeneous, that is,
$n\rightarrow n_0$, $v_x\rightarrow v_0$, $B_z\rightarrow B_0$,
$v_y\rightarrow 0$, when $x\rightarrow -\infty$.  In this case
$E_y=v_0B_0/c$. In the usual  quasi-neutrality approximation
charge-separation is weak throughout the wave profile
$
\delta n =(1/4\pi e)(dE_x/dx)\ll n.$ Further derivation is given in Appendix~\ref{perpendicular}
and results in the equation
 \begin{equation}
 \label{eq:solitoneq}
 \begin{split}
& \left(\frac{c^2}{\omega_{pe}^2}\right)\frac{1}{N}\frac{d}{dx} \frac{\gamma_e}{N} \frac{db}{dx}\\
&=  \frac{(1+\sigma)(b-1) - \sigma b(b^2-1)/2\beta_0^2}{1-\sigma(b-1)}
\end{split}
\end{equation}
where
$ N=n/n_0=v_0/v_x$, $N=(1-\sigma (b-1))(1-\sigma (b^2-1)/2\beta_0^2)^{-1}$
 and $b=B/B_0$. The obtained expression is similar to those obtained previously for nonlinear stationary waves in pair plasmas \cite{KP76}.
It is easy to see that the equations predicts  the slope scale of
$\tilde{l}_e=c\sqrt{\gamma_e}/\omega_{pe}$. The ratio
$\tilde{l}_e/r_e= (m_e/m_i)^{1/2}\sqrt{\gamma_e}\sigma^{-1/2}\ll
1$ for typical parameters of gamma-ray bursts. Therefore,
electrons are expected to be demagnetized. It has to be
understood, however, that the above small scale requires
corresponding electron drift along the shock normal to ensure the
current necessary to sustain the slope. Trajectories of
demagnetized electrons are straightened along the shock normal and
their drift is substantially suppressed, so that the ramp
steepening does not proceed to scales much smaller than those
required by the demagnetization condition. From the expression for $N$ and Eq.
\eqref{eq:solitoneq} one can see that the amplitude of the magnetic
compression reaches the values $b\sim 1/\sqrt{\sigma}$  for strongly nonlinear structures in a low-$\sigma$ plasma, in agreement with the
estimates made independently earlier in this paper.

To summarize, the basic points are the following: a) electrons
become demagnetized if the typical inhomogeneity scale becomes
smaller than the electron convective gyroradius
$c\sqrt{\gamma_e}/\omega_{pe}$ or the cross-shock electric field
slope is sufficiently steep to satisfy \eqref{eq:cond}, whichever
happens first; b) the cross-shock electric field $E_x$ is related to
the magnetic field as in \eqref{eq:phisto}, so that the potential
increases with $B^2$; c) the magnetic field, and hence the cross
shock potential, increase to high values because magnetic pressure
has to compensate the decrease of ion $T_{xx}$ as described by
\eqref{eq:eq} ); d)  large-amplitude magnetosonic waves steepen down
to the scales $c\sqrt{\gamma_e}/\omega_{pe}$,
 as described by  \eqref{eq:solitoneq}, which follows directly from the
 assumptions of electron drift and
 quasineutrality; f)  the magnetic field in these structures  increases up to
 $B/B_u\sim 1/\sqrt{\sigma}$ before the singularity $v_x=0$ is reached;
 g) according to \eqref{eq:phisto}
 the cross-shock potential is a substantial part
 of the incident ion energy, and h) the estimates above show that electrons
 have to be
 demagnetized (width is less than their convective gyroradius or
 \eqref{eq:cond} is
 satisfied). While not constituting a rigorous proof, these arguments
  show the plausibility and self-consistency of the proposed scenario of electron
 demagnetization
 by inhomogeneous cross-shock electric field and consequent heating.
 While the above scenario is described in terms of  a monotonic magnetic
 field  and potential increase across the ramp, it is likely
  that in real shocks the ramp itself breaks into sub-structures and the
  electron heating occurs as a  series of electric spikes \cite{subs}

\section{Non-magnetized shocks.}
Non-magnetized shocks are characterized by a very weak (or zero) upstream magnetic field so that the upstream convective gyroradii of both species exceed the system size and coherent magnetic braking is impossible. Weibel instability \cite{weibel} produces magnetic filaments ahead of the main transition \cite{spitk2007a}. Strong electron  heating appears to be necessary for  Weibel
mediation at $\sigma \le \eta(T_e/m_ic^2)^3$ \citep{LE2006} where
$\eta $ is a dimensionless number less than unity.   Otherwise,
Weibel turbulence is predicted to be rather small scale and weak,
so that ion scattering is relatively inefficient.    Small scale magnetic filaments, where the magnetic
field is aligned perpendicular to the shock normal in tubes or
sheets, scatter forward going electrons more readily than ions (as
does the perpendicular magnetic field enhancement for magnetized
shocks), even if the electrons are not fully magnetized. Any deflection reduces the speed along the shock normal, thus the inflowing  electrons are slowed relative to the inflowing ions by
the turbulent field. The structure is no longer one-dimensional and stationary so that \eqref{eq:basic1d}-\eqref{eq:basic2d} are not applicable directly and \eqref{eq:basic1}-\eqref{eq:basic2} should be used. However, assuming that electrons are scattered essentially randomly but small scale fields of the filaments, and neglecting ion scattering, one can average the equations over the perpendicular dimensions and time scales smaller than the ion transit time. Let us consider a single particle motion in the filamentary structure, taking the latter as given. The equations of motion read (for any species)
\begin{align}
\frac{d}{dt}p_x&=qE_x + q\hb{x}\cdot(\bm{v}_{tr}\times \bm{B}_{tr}), \label{eq:vxsingle}\\
\frac{d}{dt}\bm{p}_{tr}&=q\bm{E}_{tr} + qv_x(\hb{x}\times \bm{B}_{tr}), \label{eq:vtrsingle}
\end{align}
where $tr$ denotes $\perp \hb{x}$. Here we assume that $\bm{E}_{tr}$ and $\bm{B}_{tr}$ are small scale rapidly (in space and time) fluctuating fields \citep{spitk2007a}, while $E_x$ contains  a global coherent electric field also. Denoting by $\overline{\ldots}$ averaging over rapid fluctuations, we assume that $\overline{\bm{E}}_{tr}=0$,
$\overline{\bm{B}}_{tr}=0$, $\overline{\bm{v}}_{tr}=0$, but $\overline{{E}}_{x}\ne 0$, $\overline{{v}}_{x}\ne 0$,
$\overline{{E}^2}_{tr}\ne 0$, $\overline{{B}^2}_{tr}\ne 0$, and $\overline{{v}^2}_{tr}\ne 0$. In the lowest order approximation the particle flow is along $x$ and scattering can be treated perturbatively:
\begin{align}
\bm{p}_{tr}&\approx (q\bm{E}_{tr} + qv_x(\hb{x}\times \bm{B}_{tr}))\tau \label{eq:ptr}
\end{align}
where $\tau$ is a characteristic "collision" time. Approximating $v_x\approx c$, substituting \eqref{eq:ptr} into \eqref{eq:vxsingle} and averaging over rapid fluctuations, one has
\begin{align}
\overline{v}_x\frac{d}{dx}\overline{p}_x&= q\overline{E}_x +\frac{q^2\tau}{m\gamma} [\hb{x}\cdot\overline{(\bm{E}_{tr}\times \bm{B}_{tr})} - \overline{B^2}_{tr}], \label{eq:braking}
\end{align}
which is written for ions and electrons as well. Here we substituted
$(d/dt)\rightarrow \overline{v}_x(d/dx)$.

Simulations \citep{weibel} show that the generated magnetic field
patterns are advected toward the shock front at speeds intermediate
between the incoming plasma and the rest-frame plasma. In this case
the electric fields are substantially weaker than the magnetic
fields in the shock frame, so that the
$\hb{x}\cdot\overline{(\bm{E}_{tr}\times \bm{B}_{tr})}$ term can be
neglected relative to the last term which is nothing but the
magnetic braking due to filaments. We now involve the smallness of
$\tau$ expected from the Weibel instability. The fastest growing
modes have a scale length between the electron and ion inertial
length \citep{LE2006}. This means that in considering electron
scattering, which we propose as a physical origin of charge
separation, the ion scattering term  which is $\propto \tau/m_i$ is
small relative to the electron scattering term which is $\propto
\tau/m_e$. Thus, while the two terms in the right hand side of the
electron equation \eqref{eq:braking} may be  comparable for
electrons, the last term is neglected for ions. Therefore, the ion
motion is described by
\begin{align}
\overline{v}_{i,x}\frac{d}{dx}\overline{p}_{i,x}&= e\overline{E}_x
\label{eq:ionove}
\end{align}
and for $\overline{v}_{i,x}=c$ (negligible scattering of relativistic ions), one has
\begin{align}
c\Delta \overline{p}_{i,x}= -e\Delta\phi,
\quad \phi=-\int \overline{E}_x dx,\label{eq:ionove1}
\end{align}
in complete analogy with what happens to ions in a magnetized shock ramp:
ions are decelerated by the potential which builds up due to charge
separation caused by more efficient magnetic braking of electrons.

 With the same approximation, the electron energy changes as
follows: \begin{align}
\frac{d}{dt}(m_ec^2\overline{\gamma}_e)&=-e\overline{E_xv_x} -
e\overline{\bm{E}_{tr}\cdot\bm{v}_{tr}} \label{eq:eleove}\\ &\approx
-e\overline{E}_x\overline{v}_x +e^2
\frac{q\tau}{m_e\gamma_e}[\overline{E}_{tr}^2-\hb{x}\cdot\overline{(\bm{E}_{tr}\times
\bm{B}_{tr})}].\label{eq:eleove1} \end{align} Unless the last term
just happen to cancel the first term on the right hand side, the
electrons acquire energy which is of the order of the potential drop
$e\Delta\phi$. Since this is  the potential which decelerates ions,
$e\Delta \phi\sim m_i\gamma_0 c^2$,  therefore, \begin{align} \Delta
(m_ec^2\overline{\gamma}_e) \approx e\Delta \phi \sim m_i\gamma_0
c^2, \label{eq:eleove2} \end{align} so that electrons acquire energy
comparable to what the ions lose. Although we have not rigorously
proved that this cancelation is impossible, we may note that in
highly turbulent nonlinear environment the second term is likely to
be a highly erratic function of space and time and it does not seem
likely that its average would cancel the first term. That the first
term should be of significant size is based on the fact that
electrons are more easily scattered than the ions by the
electromagnetic turbulence and this naturally results in the
systematic charge separation during the early stages of a Weibel
shock. The efficiency of energy transfer is higher than in
magnetized shocks where only about a half of the potential can be
acquired by electrons. This is because the electrons remain almost
completely demagnetized throughout the whole region where ions
decelerate. Yet the electrons do not acquire all the momentum lost
by ions, because of their scattering. Part of the momentum is
transferred to the electromagnetic field. The pressure balance in
this case takes the form \begin{align} \sum \langle p_xv_x\rangle +
\frac{\overline{B^2}_{tr} + \overline{E^2}_{tr}-
\overline{E^2}_{x}}{8\pi}=\text{const} \label{eq:ovepressure}
\end{align} Simulations \citep{weibel,spitk2007a} show that
filaments are convected by plasma and merge, so that both the local
and average magnetic field density increase toward the shock
transition layer. This is consistent with \eqref{eq:ovepressure}:
when approaching the transition the ion momentum decreases, as well
as $\overline{E^2}_{tr}$ (the latter because of the growth of the
typical width of a filament), while $\overline{B^2}_{tr}$ should
increase. Similarly to what happens in magnetized shocks, magnetic
braking of ions is necessary to convert the energy of the directed
flow into thermal energy and decelerate the ion flow down to
sub-relativistic velocity. As a result, the magnetic field is
expected to achieve locally the equipartition values. This is also
the region where the electron scattering by the magnetic field
becomes strong. Once the electrons and ions completely thermalize
the magnetic pressure should drop to much lower magnitudes. A
transient region of a drastic local enhancement of small scale
magnetic field forms.

 Summarizing, all basic features found earlier in magnetized shocks
 (differential magnetic braking, buildup of a potential and electron
 acceleration along the shock, magnetic field increase to
 equipartition values, conversion of the directed flow energy into
 thermal energy) are also present in non-magnetized shocks; in the latter,
  the latter local inhomogeneous magnetic
 fields play the role of the large scale magnetic background of the
 former. Respectively, the spatial scales of the corresponding
 "ramp" and "overshoot" are different and determined by ion
 gyroradius in magnetized shocks, and by the filament merging in
 Weibel shocks.

\section{Synchrotron emission. } Having proposed that electrons acquire a substantial part of the incident ion energy due to the cross shock potential prior to entering a region of  a strong magnetic field, we can now estimate synchrotron emission from this region. The main radiating region in magnetized shocks is the overshoot, behind which the magnetic field drops to low values. The radiating region in non-magnetized shocks should include the filamentary region before and behind the magnetic density peak as well.    The below estimates are valid for magnetized and non-magnetized shocks as well.   Let a shock propagate
with the Lorentz factor $\gamma_0$ into interstellar medium with
the density $n_{ism}$ and magnetic field $B_{ism}$, with
$\sigma=B_{ism}^2/8\pi n_{ism}m_ic^2\ll1$. In the shock frame the
upstream density and magnetic field are $n_u=n_{ism}\gamma_0$,
$B_u=B_{ism}$. The electron energy
in the overshoot is a fraction of the incident ion energy, that
is, $\gamma_e=f_1\gamma_0/\mu$.
The overshoot magnetic field is
$B_o^2/8\pi=f_2 n_um_ic^2\gamma_0$. The electron density in the overshoot follows the ion density which remains of the same order as the upstream density, $n_e\sim n_u$. At the lower end of the energy
spectrum,the electrons emit synchrotron emission with the
characteristic frequency and power (in the shock frame),
respectively, $\omega_m=(eB_o/m_ec)\gamma_e^2$,
$P_m=(4/3)\sigma_Tc\gamma_e^2(B_o^2/8\pi)$, where $\sigma_T$ is
the cross-section of Thomson scattering. In the observer's frame
the characteristic frequency is $\omega_{obs}=\gamma_0\omega_m$,
and the emission from  unit perpendicular area becomes
$(dP/dS)_{obs}=\gamma_0^2 P_m N_s$, where $N_s=n_er_o$ is the invariant surface density of electrons. Here $r_o$ is the effective length of the radiating region.
The observed frequency and emission per unit perpendicular area are
\begin{align}
&\omega_{obs}= \left({8\pi e^2}/{m_e}\right)^{1/2} {n_{ism}^{1/2}\gamma_0^4f_1^2f_2^{1/2}}{\mu^{-5/2}},\\
&({dP}/{dS})_{obs}=2\sigma_Tm_ic^3 {f_1^2f_2\gamma_0^7
n_{ism}^2}{\mu^{-2}}r_o.
\end{align}
The largest uncertainty is in $r_o$ since there is no satisfactory theory of the relativistic shock structure (neither magnetized nor non-magnetized). In a magnetized shock the effective overshoot width is determined by the ion gyroradius in the enhanced magnetic field, $m_ic^2\gamma_0/eB_o$, times number of ion loops necessary for gyrophase mixing. The maximum overall length is expected to be of the order of the ion downstream gyroradius or less, that is,
$r_o\lesssim  f_3m_ic^2\gamma_0/B_u$, where $f_3$ may be substantially smaller than unity. Respectively,
$
({dP}/{dS})_{obs}\approx 10^9 \text{erg}/\text{cm}^2\text{s}\,\cdot(n_1^2/B_3) \gamma_{10}^7 f_1^2f_2f_3,
$
where we normalized with the
typical parameters for interstellar medium: $n_1\equiv n_{ism}/1\text{cm}^{-3}$, $B_3\equiv B_{ism}/ 3 \mu$G. For a typical gamma-ray burst $\gamma_0=10-30$ several hours after the burst, and  $\gamma_{10}\equiv\gamma_0/10$.
In Weibel mediates shocks the overshoot width is determined by the ion inertial length \citep{spitk2007a}. In this case the enhanced magnetic field is strongly inhomogeneous, so that the effective radiating width is $r_o=f_4(c/\omega_{pi})$, where $f_2$ and  $f_4$ together take into account the filling factor of about 10-15\%.   Simulations \cite{spitk2007a} show that in Weibel mediated shocks the peak magnetic density region is of the width of $\sim 50 (c/\omega_{pi})$, but the region where $B^2/8\pi \sim 0.1 n_um_ic^2\gamma_0$ may be by an order of magnitude larger.  The effective emission region may appear even substantially wider if the magnetic field decays as power law \citep{gruzinov}.  Modestly estimating for these shocks $f_1\sim 1$, $f_2f_4\sim 10^2$, one finds
$
({dP}/{dS})_{obs}\approx 10^6 \text{erg}/\text{cm}^2\text{s}\,\cdot n_1^{3/2} \gamma_{10}^7$.
  For the isotropic equivalent emitting area $ 10^{34}$cm$^2$ the total emitted power is
$
P\sim  10^{43} \text{erg}/\text{s}(n_1^2/B_3) \gamma_{10}^7 f_1^2f_2^2f_3
$ in the magnetized case and $P\sim  10^{40} \text{erg}/\text{s}\, n_1^{3/2} \gamma_{10}^7$ for non-magnetized shocks, emitted at the frequencies $\omega_{obs}\sim 10^{17} \text{s}^{1}\, n_1^{3/2}\gamma_{10}^4f_1^2f_2^{1/2}$.
In both magnetized and non-magnetized shocks the magnetic field behind the overshoot drops down, $B_d\sim B_o\sqrt{\sigma}$. Respectively, the radiation frequency drops by the same factor, while the emission power drops by the factor $1/\sigma$.

This radiation from a thin region of enhanced magnetic field may be a significant fraction of the total afterglow
emission. Consider the ratio of the afterglow from the magnetic
 region and from the entire downstream region. The
fraction of the proper hydrodynamic time scale, $\tau_h\sim
R/\gamma_0 c$, that an electron spends in the effective overshoot
region, is given by $ \tau_o/\tau_h$ which is $\sim f_3r_u/c\tau_h\sim
f_3m_ic^2\gamma_0/eB_{ism}R$ for the magnetized overshoot and $\sim f_4/\omega_{pi}\tau_h\sim$ for non-magnetized shocks. The ratio of the magnetic energy
density in the overshoot region to the average magnetic energy
downstream is $\sim 1/\sigma$.  Electron energies may remain
comparable due to effective turbulent collisions.  The relative
afterglow outputs from the overshoot region and downstream  is
then $\sim f_2^2\tau_o/\tau_h\sigma$, where $R\sim
cT_{obs}\gamma_0^2$, $T_{obs}$ being the observer time, so that
\begin{align}
&\frac{P_{\text{overshoot}}}{P_{\text{downstream}}}\sim \frac{f_2^2 f_3 \cdot 10^{-4}}{(B/3\,\mu\text{G}) (T_{obs}/10^5\,\text{s}) \sigma}\gg 1
\end{align}
for  a magnetized shock and
\begin{align}
&\frac{P_{\text{overshoot}}}{P_{\text{downstream}}}\sim \frac{f_2^2 f_4 \cdot 10^{-8}}{n_1^{1/2} (T_{obs}/10^5\,\text{s}) \sigma}\gg 1
\end{align}
for a non-magnetized shocks.
 For realistically low $\sigma$ the emission power from the
enhanced magnetic field region formally exceeds the emission power in the rest
of the  downstream region. And the typical frequencies are much
greater as well.

The cooling energy $\gamma_c$ is given by the condition
$ P_m(\gamma_c) r_o/c \sim m_ec^2\gamma_c$ and therefore
$
\gamma_c \sim \mu/\sigma_T\gamma_0^2r_on_{ism}f_2$
which corresponds to the cooling frequency in the observer's frame $\omega_c\sim\omega_{obs} (\mu\gamma_c/f_1\gamma_0)^2\sim   10^{18} \text{s}^{-1} (B_3^2 f_1^2/n_1f_2f_3)$ for the  magnetized shocks, and much higher for demagnetized shocks, which means that radiative cooling does not affect the described processes.

\section{Conclusions}
We have shown above that efficient electron heating in relativistic
collisionless shocks can be generated by a  cross-shock potential,
developing because of the preferential deflection of electrons by
the magnetic field. The cross-shock potential, which accelerates
electrons across the shock front, is of the order of the incident
ion energy, independently of whether the magnetic braking is caused
by a coherent (for magnetized shocks) or small scale (for Weibel
shocks) magnetic field. Deceleration of ions together with momentum
conservation eventually lead to strong enhancement of the magnetic
field in a small region of the shock front. This magnetic field
enhancement ensures final thermalization of ions and electrons.
Synchrotron emission from electrons from this enhanced magnetic
field region seems to be able to explain the observed afterglow
emission from gamma-ray bursts, within uncertainty of our knowledge
of plasma parameters there.

At scales below the ion gyroradius, the most likely scale for Weibel
turbulence, differential scattering of ions and electrons by
magnetic filaments can cause charge separation and strong electric
fields in the shock plane as well as a long the shock normal, so it
may be non-trivial to distinguish  a systematic cross-shock
potential from a purely stochastic electric field.  Nevertheless, we
suggest that a good way to test the idea of a systematic cross shock
potential is to compare the electric field patterns for simulated
pair shocks with shock simulations  having a realistic ion to
electron mass ratio. The mechanism we suggest, which is based upon
qualitatively different scattering of electrons and ions, works only
for large mass ratios. For pair shocks, on the other hand, electrons
and positrons can be separated by small scale magnetic fluctuations,
but there is no systematic charge separation along the shock normal.
If the systematic, cross shock  potential drop  for electron-ion
shocks is comparable to the stochastic component, it would
demonstrate the effect we are proposing.

\begin{acknowledgments}
The authors thank Y. Lyubarsky, A. Spitkovsky, and B. Katz for very
useful discussions. DE was partially supported by the Joan and
Robert Arnow Chair of Theoretical Astrophysics, the Israel-U.S.
Binational Science Foundation and by a Center of Excellence grant
from the Israel Science Foundation. MG and MAB acknowledge support
by Royal Society. MAB acknowledges support by PPARC. MG was
partially supported by BSF Grant no. 2006095 and ISF Grant no.
275/07.
\end{acknowledgments}

\appendix
\section{Two-fluid hydrodynamics of relativistic shocks}
\label{basic}
Basic equations of two-fluid relativistic hydrodynamics read
\begin{align}
& \pd{}{t}n_s + \pd{}{x_i} (n_sv_{s,i})=0, \label{eq:basic1}\\
& \pd{}{t}T_{s,i0}+ \pd{}{x_j}T_{s,ij}=n_sq_s(E_i +\epsilon_{ijk}v_{s,j}B_k/c), \label{eq:basic2}
\end{align}
where $s$ denotes the species, $i,j,k=1,2,3$, and $T_{i0}$ and $T_{ij}$ are the components of the energy-momentum tensor:
\begin{align}
& T_{0i}=\langle c p_i\rangle, \label{eq:t0i}\\
& T_{ij}=\langle v_ip_j \rangle. \label{eq:tij}
\end{align}
Here $\langle(\ldots)\rangle$ denotes averaging over the distribution function.

In the one-dimensional stationary case the equations reduce to
\begin{align}
& nv_x=\text{const}, \label{eq:basic1d}\\
& \pd{}{x}T_{ix}=nq(E_i +\epsilon_{ijk}v_{s,j}B_k/c). \label{eq:basic2d}
\end{align}
These equations should be completed with Maxwell equation  with
\begin{align}
& \rho=\sum_s n_sq_s, \\
& j_k=\sum_s n_sq_s v_{s,k}
\end{align}

\section{Electron demagnetization}
\label{demagnetization}

It is known \citep{mechanism} that in narrow nonrelativistic shocks electrons become demagnetized and efficiently heated due to the cross-shock potential. In order to find out whether such demagnetization is possible in relativistic shocks we reproduce the derivation of Ref.~\citep{mechanism} with relativistic corrections.
Namely, let us assume that a relativistic electron enters an inhomogeneous electric field $E_x=(dE/dx)x$, while the magnetic field inhomogeneity will be neglected.
It has been shown \citep{mechanism} that electron demagnetization occurs when two initially close trajectories diverge exponentially.
Let us consider two close orbits $x_1(t), y_1(t)$ and $x_2(t),y_2(t)$, each of which is a solution of the equations of motion
\begin{align}
& \frac{d}{dt} (mv_x\gamma) = -eE_x -ev_yB/c,\\
& \frac{d}{dt} (mv_y\gamma) = -eE_y +ev_xB/c.
\end{align}
The equations for the differences $\delta x=x_2-x_1$, $\delta y=y_2-y_1$, $\delta v_x=v_{2x}-v_{1x}$, $\delta v_y=v_{2y}-v_{1y}$ can be easily obtained taking into account that $\delta \gamma= \gamma^3(v_x\delta v_x + v_y\delta v_y)$:
\begin{align}
& \frac{d}{dt}[\gamma (1+\gamma^2v_x^2/c^2) \delta v_x + \gamma^3(v_xv_y/c^2) \delta v_y]= -\frac{e}{m} \frac{dE_x}{dx}\delta x - \Omega \delta v_y, \\
& \frac{d}{dt}[\gamma (1+\gamma^2v_y^2/c^2) \delta v_y + \gamma^3(v_xv_y/c^2) \delta v_x]=  \Omega \delta v_x,
\end{align}
where we assumed for simplicity that $B=\text{const}$. Here $\Omega=eB/mc$ and $E_y=\text{const}$. In the local approximation the obtained equations are linear equations with constant coefficients and substitution $ \delta x, \delta v_x, \delta v_y \propto \exp(\lambda t)$ gives
\begin{align}
& [\lambda^2 \gamma (1+\gamma^2v_x^2/c^2) +(e/m)(dE_x/dx)]\lambda^{-1} \delta v_x= -[\Omega + \lambda  \gamma^3(v_xv_y/c^2)]\delta v_y,\\
&\lambda \gamma (1+\gamma^2v_y^2/c^2)\delta v_y=[\Omega -\lambda  \gamma^3(v_xv_y/c^2)]\delta v_x,
\end{align}
so that eventually
\begin{equation}
\lambda^2\gamma^2(1+\gamma^2v^2/c^2)= - \gamma (1+\gamma^2v_y^2/c^2)(e/m)\frac{dE_x}{dx}-\Omega^2.
\end{equation}
The local criterion of instability would read
\begin{equation}
-(e/m)\frac{dE_x}{dx}>\Omega^2/\gamma (1+\gamma^2v_y^2/c^2).
\end{equation}
For electrons entering the shock without gyration $v_y=0$ and $v_x\approx c$, so that one gets trajectory divergence when
\begin{equation}
-\frac{e}{m} \frac{dE_x}{dx}- \frac{\Omega^2}{\gamma} >0,\label{criter22}
\end{equation}
with the divergence rate of
\begin{equation}
\lambda=\gamma^{-3/2} \left[-\frac{e}{m} \frac{dE_x}{dx}- \frac{\Omega^2}{\gamma} \right]^{1/2}.
\end{equation}
The demagnetized electrons are accelerated by the electric field $E_x$ across the magnetic field up to the point where the demagnetization condition ceases to be satisfied. At this point electrons begin to gyrated and all acquired energy is converted into their gyration energy. Beyond this point the only energy gain is due to the adiabatic conservation of the magnetic moment in the increasing magnetic field.

\section{Nonlinear waves and relativistic soliton}
\label{perpendicular}

We consider a stationary perpendicular wave, $\partial/\partial t=0$, $\partial/\partial y=\partial /\partial z=0$, in the framework of the two-fluid hydrodynamics of cold relativistic electrons and protons ($s=e,i$ for electrons and ions respectively):
\begin{align}
&m_sv_{sx} \frac{d}{dx}(\gamma_s v_{sx})=q_s(E_x+ v_{sy}B_z/c), \label{eq:vxeq}\\
&m_sv_{sx} \frac{d}{dx}(\gamma_s v_{sy})=q_s(E_y- v_{sx}B_z/c),\label{eq:vyeq}\\
&\gamma_s=(1-v_{sx}^2/c^2-v_{sy}^2/c^2)^{-1/2}, \label{eq:gameq}\\
&n_sv_{sx}=\text{const},\label{eq:nvxeq}\\
&E_y=\text{const}, \label{eq:eyconst}\\
&\frac{dB_z}{dx}=-4\pi \sum_s q_s n_s v_{sy}/c=4\pi e(n_ev_{ey}-n_iv_{iy})/c, \label{eq:dbzeq}\\
& \frac{dE_x}{dx}=4\pi \sum_s q_s n_s=4\pi e(n_i-n_e).\label{eq:dexeq}
\end{align}
It is worth mentioning that in the non-relativistic limit these equations have the solution in the form of the magnetosonic soliton \citep{theory} with the width $\sim c/\omega_{pe}$, where $\omega_{pe}^2=4\pi ne^2/m_e$ and the amplitude depending on the Mach number. It should be noted also that $\sim c/\omega_{pe}$ is the dispersion length of linear perpendicular magnetosonic waves.

Following the general principles of \citep{theory}, we are seeking for nonlinear wave solutions which are asymptotically homogeneous, that is, $n\rightarrow n_0$, $v_x\rightarrow v_0$, $B_z\rightarrow B_0$, $v_y\rightarrow 0$, when $x\rightarrow -\infty$.  In this case $E_y=v_0B_0/c$.
We shall consider weakly nonlinear waves in the sense that deviations from quasi-neutrality (charge-separation) are small throughout the wave profile
\begin{equation}
\delta n =\frac{1}{4\pi e}\frac{dE_x}{dx}\ll n. \label{eq:quasin}
\end{equation}
This assumption will be verified a posteriori.
In this case
$ n_e=n_i=n\Rightarrow v_{ix}=v_{ex}=v_x$,  and $nv_x=n_0v_0=\text{const}$.
Within this approximation we immediately get
\begin{align}
&m_e\gamma_ev_{ey}+m_i\gamma_i v_{iy}=0, \label{eq:sumvy}\\
& nv_0v_x(m_e\gamma_e+m_i\gamma_i) + \frac{B_z^2}{8\pi}=nv_0^2(m_e+m_i)\gamma_0 + \frac{B_0^2}{8\pi}, \label{eq:nv2}\\
 & nv_0(m_e\gamma_e+m_i\gamma_i) + \frac{v_0 B_0B_z}{4\pi c^2}=nv_0(m_e+m_i)\gamma_0  + \frac{v_0B_0^2}{4\pi c^2}, \label{eq:nve}\\
\end{align}
where $\gamma_0=(1-v_0^2/c^2)^{-1/2}$, and further
\begin{align}
& (m_e\gamma_e+m_i\gamma_i)= (m_e+m_i)\gamma_0 - \frac{B_0^2}{4\pi n_0c^2} (b-1), \label{eq:mgammas}\\
& \frac{v_x}{v_0}=\frac{1-\sigma (b^2-1)/2\beta_0^2}{1-\sigma (b-1)}, \label{eq:vxfinal}
\end{align}
where $\sigma =B_0^2/4\pi n_0(m_i+m_e)c^2\gamma_0$, $\beta_0=v_0/c$, and $b=B_z/b_0$. It is easy to see that $v_x$ is a monotonically decreasing function of $b$ in the range $1\leq b<1+1/\sigma$.

Using \eqref{eq:sumvy} and \eqref{eq:dbzeq} one obtains
\begin{equation}
\frac{(m_e\gamma_e+m_i\gamma_i)v_{iy}}{m_e\gamma_e}= -\frac{c}{4\pi n e} \frac{dB_z}{dx}.
\end{equation}
In what follows we
shall make the assumption that the energy content in ions is always much higher than in electrons, that is,
\begin{equation}
m_i\gamma_i \gg m_e\gamma_e, \label{eq:gammas}
\end{equation}
so that approximately
\begin{equation}
m_i\gamma_iv_{iy}=-\frac{cm_e\gamma_e}{4\pi e n} \frac{dB_z}{dx}.\label{eq:vey}
\end{equation}
Following the path outlined in the non-relativistic analysis \citep{theory} we substitute \eqref{eq:vey} into \eqref{eq:vyeq} for ions to obtain
\begin{equation}
-v_x\frac{d}{dx} \left(\frac{m_e\gamma_e}{4\pi e^2n}\right)\frac{dB_z}{dx}= v_0B_0-v_xB_z,
\end{equation}
 or, after normalization,
 \begin{equation}
 \label{eq:solitoneq1}
 \begin{split}
& \left(\frac{c^2}{\omega_{pe}^2}\right)\frac{1}{N}\frac{d}{dx} \frac{\gamma_e}{N} \frac{d}{dx} b\\
&=  \frac{(1+\sigma)(b-1) - \sigma b(b^2-1)/2\beta_0^2}{1-\sigma(b-1)}
\end{split}
\end{equation}
where $N=n/n_0=v_0/v_x$.

In the asymptotically homogeneous point  one has
\begin{equation}
\left(\frac{c^2}{\omega_{pe}^2}\right)\gamma_0\frac{d^2}{dx^2} \xi = (1+\sigma - \sigma/\beta_0^2) \xi,
\end{equation}
where $\xi =b-1\ll 1$. This point is unstable when $\beta_0^2>\sigma/(1+\sigma)$, in which case \citep{theory} the solution should be of a soliton type (non-periodic wave).  The electron Lorentz-factor $\gamma_e$  cannot be represented as a function of $b$, so that  \eqref{eq:solitoneq1} cannot be converted to a quasi-potential equation . However, we can use the fact that $\gamma_e>0$ to define a new coordinate $dw=Ndx/\gamma_e$, so that
\begin{equation}
\left(\frac{c^2}{\omega_{pe}^2}\right) \frac{d^2}{dw^2}b=\gamma_e \frac{(1+\sigma)(b-1) - \sigma b(b^2-1)/2\beta_0^2}{1-\sigma(b-1)}.\label{eq:derived}
\end{equation}
The derived equation is valid provided that the flow does not come to a halt, $v_x>0$, that is,
\begin{equation}
b<b_c=\sqrt{1+2\beta_0^2/\sigma}<1+1/\sigma.  \label{eq:nondivergent}
\end{equation}

When $b$ increases the right hand side remains positive until $b(b-1)=2\beta_0^2(1+\sigma)/\sigma$. For $\sigma\ll 1$ and $\gamma_0\gg 1$ this means that the sign changes when $b=\sqrt{2/\sigma}\gg 1$. At this point the denominator $1-\sigma b\approx 1$. It is well-known \citep{KP76} that there are no soliton solutions for $\sigma\ll 1$ in the pair plasma, where $\gamma_e=\gamma_i=1-\sigma(b-1)$. For a soliton solution to exist
\begin{equation}
\int_1^{b_m} \gamma_e \frac{(1+\sigma)(b-1) - \sigma b(b^2-1)/2\beta_0^2}{1-\sigma(b-1)}db=0
\end{equation}
has to be satisfied for $b_m<b_c$. Although complete analysis is impossible here  it is likely that a soliton solution would not exist for too low $\sigma$ for the electron-ion plasma as well.

For the analysis of the solution behavior it is sufficient to know that $\gamma_0 \leq \gamma_e\lesssim \gamma_i(m_i/m_e)$. It is easy to estimate the typical inhomogeneity scalce as $l_s\sim (c/\omega_{pe}) \gamma_e^{1/2}$. For $\sigma\ll 1$ (typical for gamma-ray bursts) the highest achievable magnetic field amplitude would grow as $b_{max}\sim 1/\sigma^{1/2}$, thus ensuring strong magnetic compression. Since $\sigma b\ll 1$ always, the electron current can be estimated as follows:
\begin{equation}
nev_{ey}\sim \frac{n_0ec}{1-\sigma b^2/2},
\end{equation}
where we assume that electrons remain relativistic: since $v_x$ becomes sub-relativistic, $v_{ey}\sim c$. Then the typical length of the magnetic field variation is
\begin{equation}
\left|\frac{B}{(dB/dx)}\right|\sim \frac{B_0b(1-\sigma b^2/2) }{4\pi en_0}\sim \frac{c}{\omega_{pe}} (M\sigma)^{1/2} b(1-\sigma b^2/2)
\end{equation}
where $M=m_i/m_e$. The maximum length is achieved when $b\sim 1/\sqrt{\sigma}$ and $1-\sigma b^2/2\sim 1$, where $l\sim c/\omega_{pi}$. For smaller $b\sim a/\sqrt{\sigma}$, $a\ll 1$, the length $l\sim a (c/\omega_{pi})$, while for highest possible $b\sim 1/\sqrt{\sigma}$ and $1-\sigma b^2/2\sim \sqrt{\sigma}$, and  the length becomes $l\sim (c/\omega_{pe}) (M\sigma)^{1/2}$.

It has to be understood, however,  that  the obtained expressions provide only an indication of the character of the wave steepening. Indeed, strong magnetic compression and narrow width ensure that ions behave nonadiabatically and begin to gyrate strongly in the vicinity of the magnetic field maximum. The ion gyration makes the cold hydrodynamical approximation invalid. Thus, the derived equation \eqref{eq:solitoneq1} provides a satisfactory estimate of the wave profile only at the upstream edge of the shock ramp \citep{theory}, which nevertheless is quite sufficient for physical conclusions to be made.

Using \eqref{eq:vxeq} one can find
\begin{align}
\tfrac{1}{2}\frac{d}{dx} [(m_i^2\gamma_i^2+m_e^2\gamma_e^2)v_x^2]= -e(m_i\gamma_i+m_e\gamma_e)\frac{d\varphi}{dx}.
\end{align}
Taking into account the above approximation $m_e\gamma_e\ll m_i \gamma_i$,  and the expressions \eqref{eq:mgammas} and \eqref{eq:vxfinal}, one gets
\begin{equation}
\frac{e\varphi}{m_ic^2\gamma_0}= \int_{b_0}^b \left(\frac{\sigma b}{\beta_0^2}\right) \left(\frac{1-\sigma(b^2-1)/2\beta_0^2}{1 -\sigma (b-1)}\right) db.
\end{equation}
For the above approximation the potential from the asymptotically homogeneous point to the point where $d^2b/dw^2=0$ is easily evaluated as $e\phi\approx 0.5m_ic^2\gamma_0$.


\end{document}